\documentclass[preprint,showpacs,preprintnumbers,amsmath,amssymb]{revtex4}

\usepackage{graphicx}
\usepackage{dcolumn}
\usepackage{bm}

\begin{document}

\voffset=0.5truein

\preprint{LA-UR-03-0239}

\title{Thermal Stabilization of the HCP Phase in Titanium}

\author{Sven P. Rudin$^1$}
\author{M. D. Jones$^2$}
\author{ R. C. Albers$^1$}
\affiliation{$^1$Los Alamos National Laboratory, Los Alamos, NM 87545 \\
$^2$Department of Physics and
Center for Computational Research, University at Buffalo,
The State University of New York, Buffalo, NY 14260
}

\date{\today}

\begin{abstract}
We have used a tight-binding model that is fit to first-principles
electronic-structure calculations for titanium to calculate
quasi-harmonic phonons and the Gibbs free energy of
the hexagonal close-packed (hcp) and omega ($\omega$)
crystal structures.
We show that the true zero-temperature ground-state is
the $\omega$ structure, although this has never been observed
experimentally at normal pressure, and that it is
the entropy from the thermal population
of phonon states which stabilizes the hcp structure
at room temperature.
We present the first completely theoretical prediction of
the temperature- and pressure-dependence of the
hcp-$\omega$ phase transformation and show that
it is in good agreement with experiment.
The quasi-harmonic approximation fails to adequately
treat the bcc phase because the zero-temperature phonons of
this structure are not all stable.
\end{abstract}

\pacs{63.20.Dj, 64.30.+t, 64.70.Kb, 65.40.Gr}
\maketitle


The experimentally observed ground state of titanium
at room temperature is the
hexagonal close-packed (hcp) structure\cite{sikka82},
even though zero-temperature first-principles electronic-structure
calculations\cite{gyanchandani90, ahuja93, jomard98, greeff01, mehl02}
predict the omega ($\omega$) structure
to be the energetically favored ground state.
There are two possible causes for this discrepancy:
(1) this is another failure of the local-density approximation (LDA) for
calculating electronic-structure of solids, or (2) thermal effects
stabilize the hcp phase.  In this paper we show that
the second argument is correct: the
thermal occupation of phonon states favors the hcp
structure above a pressure-dependent transition temperature.
Only by including the vibrational entropy in the first-principles
calculations can theory be brought into agreement with experiment.

Titanium is a good case for testing our ability to predict
solid-solid phase transformations with first-principles methods.  It is
an element that displays a rich phase diagram with several recently
discovered high-pressure phases\cite{vohra01,akahama01}.
When alloyed with other elements, it leads to materials of major
technical importance for the airline, space, and other industries.

It is possible to use first-principles methods
to calculate the zero-temperature internal energy, $\Phi_0 (V)$, 
as well as the free energy contributions from the ions,
$F_I (T,V)$, and the electrons, $F_E (T,V)$, to give the complete
equation of state
\begin{equation}
\label{eq:freeen1}
F (T,V) = \Phi_0 (V) + F_I (T,V) + F_E (T,V).
\end{equation}
Because of the large computational effort required to calculate
the free energy of Eq.~\ref{eq:freeen1}
over a significant range of volumes and temperatures,
we have chosen to use a tight-binding model instead
of conventional first-principles electronic-structure methods.
We will demonstrate the high accuracy of our model and
believe that it should be essentially identical with the
best available first-principles methods.

\begin{figure}
\includegraphics[width=8cm]{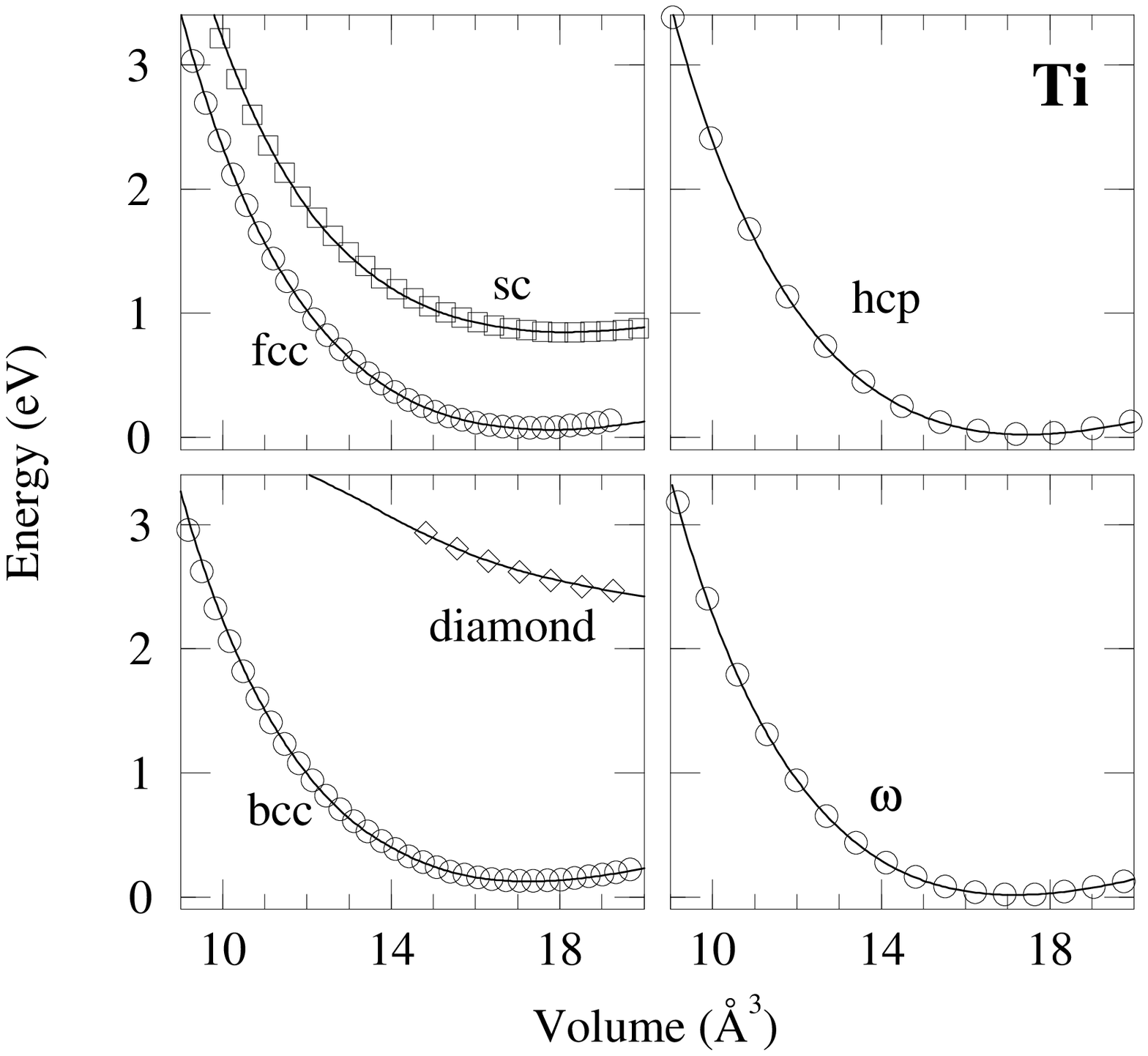}
\caption{\label{fig:fitdata}
Calculated energies for the crystal structures in the fitting database.
Symbols are the first-principles results, solid lines are calculated from
the fitted tight-binding model.
Also included in the fit (but not shown here) are
the fcc crystal with distortions corresponding to the
longitudinal and transverse phonons at the reciprocal-space high-symmetry
point X.
The $\omega$ structure lies lowest in energy,
in contrast to the experimentally found ground state, hcp.
}
\end{figure}

The model's accurate emulation of first-principles calculations
(in this case the Wien97 code\cite{wien97}) requires
optimization of a single tight-binding parameterization such that the model
closely reproduces the relative total energies of selected
crystal structures.
The structures are chosen to be both simple and representative:
simple enough to make the first-principles calculations feasible,
and representative of all microscopic environments
the model is aimed at describing.
A comparison of total energies as a function of volume with
structures fulfilling these requirements are shown in
Fig.~\ref{fig:fitdata}.
Not shown but included in the fit as independent structures are
fcc crystals distorted to correspond to the
longitudinal and transverse phonons at the reciprocal-space high-symmetry
point X. These additional structures
fine-tune the model's ability to correctly
describe lattice vibrations\cite{rudin02}.
Crucial for convergence of the fitting procedure is
to fit the cubic-structure energy bands at high-symmetry points
in reciprocal space\cite{mat02,handbook}.

The electronic contribution to the free energy
depends on the electronic density of states (DOS)
as a function of volume, $n(E,V)$.
The occupation of these states,
given by the Fermi distribution $f(E,T) =
[e^{(E-E_f)/(k_BT)}+1]^{-1}$,
determines their entropy\cite{pathria},
\begin{equation}
\label{eq:EntElec}
S_{el} (T,V) = -k_B \int [f \ln f + (1-f) \ln(1-f)] n(E,V) dE
,\end{equation}
and hence the electronic contribution
to the free energy $F_E(T,V) = -T S_{el} (T,V)$.
This contribution is small compared to that of the ions but is
included for completeness.

The ionic contribution to the free energy depends on the phonon DOS,
$g(\omega,V)$, through the
zero-point energy,
\begin{equation}
\label{eq:ZPE}
U_{zero} (V) = \frac{1}{2} \int_\Omega \hbar\omega g(\omega,V) d\omega
,
\end{equation}
as well as the temperature-dependent,
quasi-harmonic free energy,\cite{pathria}
\begin{equation}
\label{eq:FEPhon}
F_{H} (T,V)
= k_B T \int_\Omega \ln[1-e^{-\hbar\omega/k_BT}] g(\omega,V) d\omega
.\end{equation}
This differs from the volume-independent harmonic approximation
in that the quasi-harmonic approximation uses phonons calculated
for small displacements (i.e., harmonic) at each volume.
The resulting phonon frequencies are therefore volume dependent
but temperature independent (they are calculated at
zero temperature).
The approximation requires that the phonons for any given crystal
structure and volume
are stable at zero temperature, which may not be true if the crystal
can lower its total energy by spontaneously and continuously distorting
into another crystal structure. 
At elevated temperatures (near melting) this approximation also
neglects some important anharmonic contributions to the free energy
that can be responsible for stabilizing some
high-temperature phases.

We have used the tight-binding model to calculate
the electronic and the phonon DOS
by evaluating the relevant energy eigenvalues and dynamical matrices
on a fine mesh of wave vectors in the first Brillouin zone.
In both cases the spectrum is smeared with a Gaussian, chosen to be
as small as possible while keeping the DOS smooth and continuous.
The mesh is refined until the free energy converges.

The dynamical matrix at a given wave vector {\bf q} is the Fourier
transform of the force constants, which we calculate
from the tight-binding model by the
direct-force method\cite{kunc82,wei92,frank95,parlinski97}.
This method requires large simulation cells consisting of
repeated unit cells transposed by vectors ${\bf \ell}$,
the forces on all atoms are calculated in response to the
displacement of the atom(s) in one unit cell.
Our simulation cells contain
54 (hcp), 81 ($\omega$), and 128 atoms (bcc);
for more details we refer the reader to Ref.\cite{rudin02}.

\begin{figure}
\includegraphics[width=8cm]{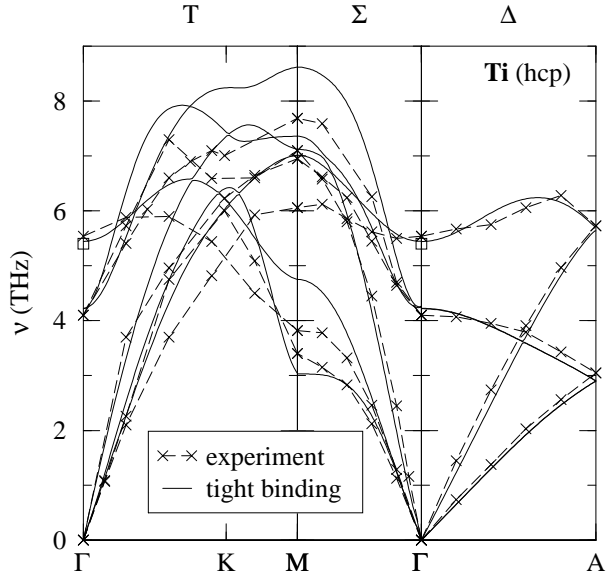}
\caption{\label{fig:disper}
Experimental (crosses)\cite{stassis79} and tight-binding (solid lines) phonon
dispersion of titanium in the hcp crystal structure at ambient
pressure.
Although some experimental details are not reproduced by
our tight-binding model, the overall agreement is quite good.
The phonon DOS  that is
relevant for the calculation of the Gibbs free energy
should still be highly accurate.
The square at $\Gamma$ is from a first-principles
frozen-phonon calculation.
}
\end{figure}

Figure~\ref{fig:disper}
compares our calculated phonon dispersion for the hcp structure
to experimental data\cite{stassis79}.
The overall agreement is good, in particular near the zone center
(around $\Gamma$) and along the vertical direction (parallel to the
c-axis).
Frequencies of modes in horizontal directions (perpendicular to
the c-axis) away from the zone center are somewhat high compared to experiment.
The main characteristics, which the tight-binding model captures,
give a reliable DOS, the relevant entity used in
Eqs.~\ref{eq:ZPE}~and~\ref{eq:FEPhon}
to calculate the phonon contribution to the free energy.

The phonon density of states is calculated for the
hcp\footnote{The $c$ over $a$ ratio of the hcp structure is varied to give the
lowest zero-temperature energy, i.e., in particular at high pressures
the calculation is done for a hexagonal crystal with a $c/a$ 
larger than the ideal value.},
$\omega$, and
bcc structures at a dozen volumes corresponding to pressures in the range
from zero to 170~GPa and is then used to
evaluate the Gibbs free energy\cite{pathria},
$G(T,P) = F(T,V)+PV$.

\begin{figure}
\includegraphics[width=8cm]{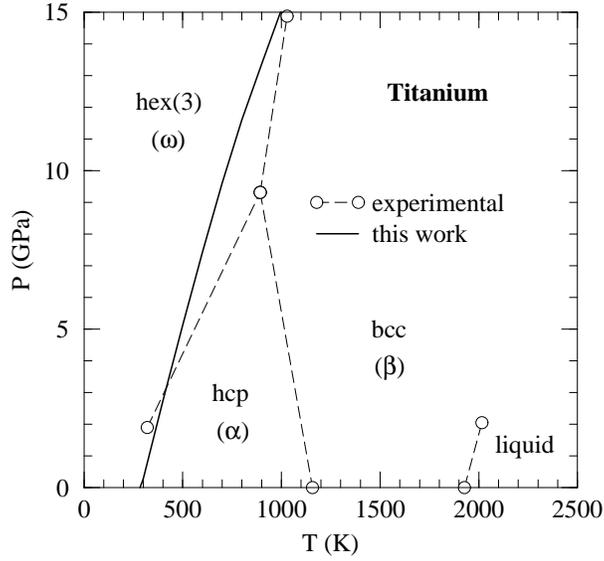}
\caption{\label{fig:phasedia}
The phase diagram of titanium.
The dashed lines connect the
experimental data points given by Young\cite{young91};
the solid line shows our calculated
$\alpha\rightarrow\omega$
transformation.
Thermal stabilization explains why
at room temperature the experimentally observed hcp structure
is favored over the $\omega$ structure
(the calculated zero-temperature ground state).
}
\end{figure}

Comparison of the Gibbs free energies as a function of pressure and
temperature for the hcp and $\omega$ crystal structures
allows us to identify the phase transformation line.
This approach is extremely demanding because tiny errors in the relative
free energies can dramatically alter the phase boundary.
Figure~\ref{fig:phasedia} shows good agreement between our
theoretically calculated $\alpha\rightarrow\omega$ phase transformation
line with the experimental phase diagram\cite{young91}.
The calculated transition temperature at zero pressure is
approximately 280~K.

At normal pressure we know of no experimental data that shows
that the $\omega$ phase is the correct zero-temperature ground state.
However, it has recently been shown that energy barriers exist for this phase
transformation\cite{trinkle03}  that may make it extremely difficult to
verify at low temperatures.  Thus the kinetics may be just too sluggish for 
the
system to bring itself into true thermodynamic equilibrium.

We were not able to similarly calculate and include in Fig.~\ref{fig:phasedia}
the  transition into the bcc structure.
The open bcc structure appears to be
dynamically stabilized by entropy and anharmonic effects.
At zero temperature some phonon modes are unstable
due to a spontaneous instability of the bcc crystal structure into
the $\omega$ phase.  There is no energy barrier to this
continuous transformation.
In particular, this causes the longitudinal phonon mode with wave vector
${\bf q}=\frac{2}{3}(1,1,1)$, which distorts the bcc crystal into the
$\omega$ phase, to be unstable below pressures of roughly
40~GPa.
The mechanical instability of this phonon mode at low temperatures
is also seen experimentally; the martensitic
phase transition from the bcc phase to the hcp phase
prevents the high-temperature
bcc phase from being quenched to room temperature\cite{petry91}.
One branch of transverse phonon modes with wave vectors $(\xi,\xi,0)$,
which are related to the bcc to hcp transition, also appears
unstable in our treatment.
Both of these instabilities are consistent with strictly first-principles
calculations of zero-temperature phonons\cite{persson00}.

\begin{figure}
\includegraphics[width=8cm]{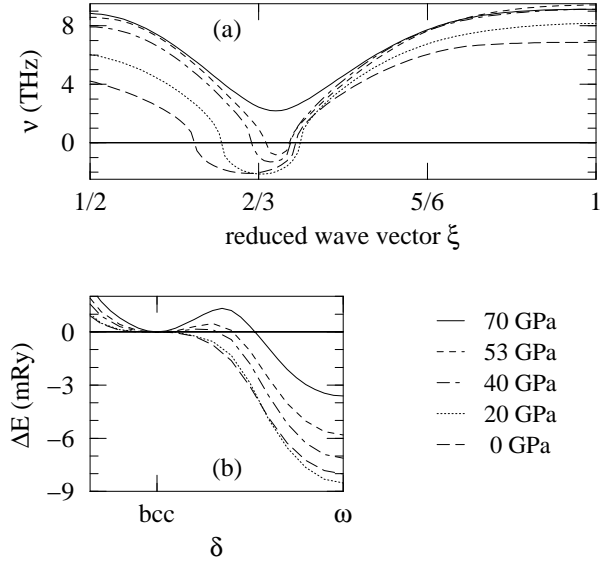}
\caption{\label{fig:bccbranch}
(a) Calculated zero-temperature
longitudinal phonon modes around the wave vector
${\bf q}=\frac{2}{3}(1,1,1)$, which distorts the bcc crystal into the
$\omega$ phase.
(b) Calculated zero-temperature
energy landscape for the bcc to $\omega$ pathway.
}
\end{figure}

Figure~\ref{fig:bccbranch}a shows
the relevant
segment of the $(\xi,\xi,\xi)$ branch.
The energy of the corresponding
pathway from the bcc phase to the $\omega$ phase,
shown in Fig.~\ref{fig:bccbranch}b, develops a local minima for
the bcc phase at pressures of roughly 40~GPa, but at lower pressures
the crystal can deform from bcc to $\omega$ without encountering an
energy barrier.
This pathway can be viewed as planes moving in accordance with the
wave vector ${\bf q}=\frac{2}{3}(1,1,1)$ or as chains of atoms along
the body diagonal moving with respect to each other.
Calculations of the energy needed to move such a single chain of
atoms indicate that for small displacements the movement is energetically
favored, whereas the movement of a single atom is not.

If only the stable phonons are included in the evaluation of
Eq.~\ref{eq:FEPhon},
the bcc structure is calculated to be lowest in Gibbs free energy at
pressures that are an order of magnitude higher than
experimentally observed.
A correct calculation
requires a more sophisticated treatment that includes
phonon-phonon interactions, i.e., anharmonic contributions to the
free energy.

In conclusion, we have for the first time used a completely
theoretical approach, without any experimental input, to
calculate the Gibbs free energies of competing crystal phases in titanium.
Comparison of the hcp and $\omega$
Gibbs free energies results in a pressure- and temperature-dependence
of the structural
phase transition in good agreement with experiment and serves to explain
how finite-temperature entropy effects stabilize
the experimentally found room-temperature hcp crystal structure
over the theoretically found zero-temperature $\omega$-phase ground state.

We thank Matthias Graf, Carl Greef, and Dallas Trinkle
for helpful and encouraging discussions.
This research is supported by the Department of Energy under contract
W-7405-ENG-36. All FLAPW calculations were performed using the Wien97
package\cite{wien97}.

\bibliography{backbone}

\end{document}